\documentclass[12pt]{iopart}

\usepackage{graphicx}
\usepackage[usenames]{color}

\begin{document}


\title{Stable and mobile  excited
two-dimensional dipolar  Bose-Einstein condensate
 solitons}

\author{ Adhikari S K \footnote{Adhikari@ift.unesp.br; URL: http://www.ift.unesp.br/users/Adhikari}
} 
\address{
Instituto de F\'{\i}sica Te\'orica, UNESP - Universidade Estadual Paulista, 01.140-070 S\~ao Paulo, S\~ao Paulo, Brazil
} 

\begin{abstract}

We demonstrate 
robust, stable, mobile excited states of    
quasi-two-dimensional (quasi-2D) dipolar Bose-Einstein condensate (BEC) solitons     
   for repulsive  
contact interaction with a harmonic   trap along the $x$ direction perpendicular to the polarization direction $z$. Such a soliton can freely move in the $y-z$ plane.
A rich variety of such excitations is considered:
one quanta of excitation for movement along (i) $y$ axis or  (ii) $z$ axis or 
(iii) both.   A proposal for creating these excited solitonic states  in a
laboratory by phase imprinting  
is also discussed. We also consider excited states of quasi-2D dipolar BEC soliton 
where the sign of the   dipolar interaction is reversed by a rotating orienting field. 
In this sign-changed case the soliton moves freely in the $x-y$ plane under the action of a harmonic trap in the $z$ direction. 
At medium  velocity the head-on collision of two such solitons is found to be quasi elastic with practically no deformation. 
  The findings are illustrated using numerical simulation in three and two spatial  dimensions employing realistic   interaction parameters for a dipolar $^{164}$Dy BEC.

\end{abstract}


\maketitle

\section{Introduction}
 
A bright soliton is a localized intensity peak 
that maintains its shape, while
traveling at a constant velocity in one dimension (1D), due to a cancellation of nonlinear attraction and dispersive
effects \cite{rmp}.  Solitons have been observed  in water waves, nonlinear optics, and Bose-Einstein condensate (BEC) etc. among others. 
In physical three-dimensional (3D) world, quasi-one-dimensional (quasi-1D) solitons are observed where a reduced (integrated) 1D
density exhibit soliton-like property \cite{4}.  
Experimentally, bright matter-wave solitons and soliton trains were created in a BEC of
$^7$Li \cite{1}
 and
$^{85}$Rb atoms \cite{3}.
However, due to collapse instability, 3D BEC
bright solitons  are fragile and can accommodate only a small 
number of atoms \cite{4}.

Recent 
observation of  BECs of $^{164}$Dy \cite{ExpDy,dy}, $^{168}$Er \cite{ExpEr} and 
  $^{52}$Cr \cite{cr,saddle} atoms
with large magnetic dipole
moments has opened new directions of research in BEC solitons.
In a dipolar BEC, in addition to the quasi-1D solitons \cite{1D} it is possible to have a quasi-two-dimensional (quasi-2D) soliton \cite{2D} free to move in a plane with a constant velocity and trapped in the perpendicular direction. Although, 
such a quasi-2D dipolar BEC soliton has not been experimentally observed, this seems to be 
the only simplest example of an experimentally realizable  quasi-2D soliton.
Dipolar BEC solitons can be stabilized either by a harmonic or an
optical-lattice trap
 in quasi-1D \cite{ol1D} and quasi-2D \cite{ol2D} configurations.
More interestingly,  one can have dipolar BEC solitons for fully repulsive 
contact interaction \cite{1D}.  Hence these dipolar 
solitons bound by long-range dipolar  interaction could be  robust and less vulnerable to collapse for a large number of atoms due to the 
repulsive contact interaction. The contact repulsion gives stability against collapse and 
dipolar attraction prevents the escape of the atoms from the soliton.

The possibility of realizing a trapped BEC in an excited (non-ground) state 
has been a topic of intense research \cite{excited}. Here we demonstrate the  existence of stable excited quasi-2D dipolar BEC solitons harmonically trapped in the $x$ direction 
capable of moving in the $y-z$ plane with a constant velocity with $z$ the polarization direction. 
  They are stable and  stationary excitations
of the quasi-2D bright solitons. 
  We consider three types of excitations of the solitons: lowest excitation for dynamics along 
(i) $y$ axis, (ii) $z$ axis, and (iii) both $y$ and $z$ axes. The wave function for the lowest $y$ ($z$) excitation is antisymmetric in $y$ ($z$) with a zero in the  $y=0$ ($z=0$) plane.     The symmetries of these states are quite similar to the excited states of a 3D trapped harmonic oscillator, the only difference is that in the case of the quasi-2D dipolar solitons  there is no harmonic trap along $y$ and $z$ axes and the confinement in these directions is achieved solely by dipolar attraction.     

In addition to the normal excited  quasi-2D dipolar BEC soliton, we also considered a set-up where the sign of the dipolar interaction is reversed by a rotating orienting 
field \cite{orienting}.  In this sign changed configuration, the dipolar interaction is repulsive along $z$ axis and attractive in the $x-y$ plane and an excited  quasi-2D dipolar BEC soliton can be realized by applying a harmonic trap along the $z$ axis. The stationary 
excited quasi-2D bright solitons were obtained by a imaginary-time propagation of the mean-field Gross-Pitaevskii (GP) equation  
in 3D. These stationary excited solitons can also be termed quasi-2D dipolar dark-in-bright solitons 
bearing resemblance with the quasi-1D dipolar dark-in-bright solitons studied recently \cite{dib1d}. The quasi-1D dipolar dark-in-bright solitons were also established to be excitations of 
dipolar bright solitons \cite{dib1d,black}.

We also studied the dynamics of these quasi-2D solitons by real-time simulation using an effective 2D mean-field model.   The stability of the excited quasi-2D dipolar solitons was established 
by studying the breathing oscillation of the system upon small perturbation over long time in real-time propagation.  
The head-on collision between two excited quasi-2D solitons 
 is found to be quasi elastic at medium  velocities of few mm/s. 
In such a collision, two excited solitons  pass through each other without significant deformation.  
However, at lower velocities the collision becomes inelastic, as only the strictly 1D integrable solitons can have elastic collision at all velocities \cite{rmp}.   We also demonstrate by real-time simulation in 2D
the possibility of the creation of the excited quasi-2D solitons by phase imprinting \cite{darksol,phase} a normal  quasi-2D bright soliton with identical parameters with, for example, a phase difference of $\pi$ between the  two parts of the BEC wave function situated at $y>0$ and $y<0$.

In Sec. II the time-dependent 3D mean-field model for a quasi-2D 
dipolar BEC soliton is presented. A reduced 2D model appropriate for the present study 
is also considered. The numerical investigation of the quasi-2D solitons is considered in 
Sec. III. The domain of the appearance of the ground and excited states of the  quasi-2D solitons 
is illustrated in a phase plot with realistic values of contact and dipolar interactions of $^{164}$Dy atoms exhibiting
the maximum number of atoms in the quasi-2D soliton
versus the scattering length. The  evolution of collision between  two excited quasi-2D  
solitons is considered by real-time evolution. The dynamical simulation of the creation of an excited  quasi-2D soliton 
from a phase-imprinted ground state of a quasi-2D soliton is also demonstrated. 
The possibility of creating a dark soliton by phase imprinting a BEC by means of a detuned  
laser has been illustrated experimentally \cite{darksol}.
Finally, in Sec. IV we present a brief summary and concluding remarks.

\section{Mean-field Gross-Pitaevskii equations}

We consider a dipolar BEC soliton, with the 
mass, number of atoms, magnetic  dipole moment, and scattering length 
given by $m, N, 
\mu, a$.   The 
interaction between 
two atoms at  $\bf r$ and $\bf r'$ is  \cite{rpp}
\begin{eqnarray}\label{intrapot} 
V_i({\bf R})= 3\alpha
a_{\mathrm {dd}}V_{\mathrm {dd}}({\mathbf R})+4\pi a \delta({\bf R})
, \quad \bf R = (r-r'),
     \end{eqnarray}
 with \begin{eqnarray}  a_{\mathrm {dd}}=
\frac{\mu_0  \mu^2m}{12\pi \hbar ^2    }, \quad
V_{\mathrm {dd}}({\mathbf R})=\frac{1-3\cos^2 \theta}{{\bf R}^3},
\end{eqnarray}where 
  $\mu_0$ is the permeability of free space, 
$\theta$ is the angle made by the vector ${\bf R}$ with the polarization 
$z$ direction. The parameter $\alpha$ ($1>\alpha>-1/2$) can be tuned by a rapidly rotating magnetic field 
allowing the change of the sign of dipolar interaction \cite{orienting}. We will 
consider two cases in this study: (i) $\alpha =1$ corresponding to the normal dipolar interaction and (ii) $\alpha = -1/2$    corresponding to a sign-changed dipolar interaction. 

The first 
possibility (i) leads to {\it a fully asymmetric} quasi-2D excited solitonic state 
 in the presence of  a harmonic trap along $x$ axis.
The dimensionless GP equation in this case    
can be written as   \cite{rpp}
\begin{eqnarray}&& \,
{ i} \frac{\partial \phi({\bf r},t)}{\partial t}=
{\Big [}  -\frac{\nabla^2}{2 }
+
\frac{1}{2} x^2
\nonumber \\  &&  \,
+ g \vert \phi \vert^2
+ g_{\mathrm {dd}}
\int V_{\mathrm {dd}}({\mathbf R})\vert\phi({\mathbf r'},t)
\vert^2 d{\mathbf r}'  
{\Big ]}  \phi({\bf r},t),
\label{eq3}
\end{eqnarray}
where 
$g=4\pi a N,$ 
$g_{\mathrm {dd}}= 3N a_{\mathrm {dd}}$. 
In  (\ref{eq3}), length is expressed in units of 
oscillator length  $l=\sqrt{\hbar/(m\omega)}$,  where $\omega$  is the circular frequency 
of the harmonic trap acting along  $x$ axis. The 
energy is in units of oscillator energy  $\hbar\omega$, probability density 
$|\phi|^2$ in units of $l^{-3}$, and time in units of $ 
t_0=1/\omega$. 

The second possibility (ii) leads to {\it an axially-symmetric} excited quasi-2D dipolar soliton in the presence of  a harmonic trap  along $z$ axis. The dimensionless GP equation in this case is \cite{2D}
\begin{eqnarray}&& \,
{ i} \frac{\partial \phi({\bf r},t)}{\partial t}=
{\Big [}  -\frac{\nabla^2}{2 }
+
\frac{1}{2} z^2
\nonumber \\  &&  \,
+ g \vert \phi \vert^2
-\frac{1}{2} g_{\mathrm {dd}}
\int V_{\mathrm {dd}}({\mathbf R})\vert\phi({\mathbf r'},t)
\vert^2 d{\mathbf r}'  
{\Big ]}  \phi({\bf r},t),
\label{eq4}
\end{eqnarray}
where now $\omega$ is the circular frequency of the harmonic trap along $z$ axis. 
In the case of  (\ref{eq3}) the dipolar interaction is attractive along $z$ axis and repulsive in the $x-y$ plane;
the opposite is true for  (\ref{eq4}). The anisotropic dipolar interaction is circularly symmetric in the $x-y$ plane in 
both cases.
Hence,  (\ref{eq3}) is fully anisotropic and leads to fully-anisotropic quasi-2D solitons, whereas  (\ref{eq4}) is 
axially symmetric and hence leads to quasi-2D solitons with this symmetry. 
 
In place of  (\ref{eq4}), a quasi-2D model appropriate for the 
quasi-2D excited solitons of large spatial extension  is very economic and convenient 
from a computational point of view, specially for real-time dynamics. 
The system is assumed to be in the ground state $\phi(z)=\exp (-z^2/2)/(\pi )^{1/4}$
of the axial  trap and the wave function can be written as 
$\phi({\bf r},t)=\phi(z) \phi_{2D}({\vec \rho},t)$, where 
$ \phi_{2D}({\vec \rho},t)$ is an effective wave function in the $x-y$ plane. Using this ansatz in  (\ref{eq4}), the $z$ dependence can be integrated out to obtain the following effective 2D equation \cite{laser2D}
\begin{eqnarray}&&
{ i} \frac{\partial \phi_{2D}({\vec \rho},t)}{\partial t}=
\biggr[ - \frac{\nabla_\rho^2}{2}+ \frac{g}{\sqrt{2\pi}}|\phi_{2D}
(\vec \rho,t)|^2  \label{eq5}
\nonumber \\
&&+\frac{4\pi g_{dd}}{3\sqrt{2\pi}}\int \frac{d{\bf k_\rho}}{(2\pi)^2}
e^{i{\bf k_\rho}\cdot {\vec \rho}} n({\bf k_\rho},t)h_{2D}\left(\frac{k_\rho}{2}\right)\biggr] \phi_{2D}(\vec \rho,t), \\
&&n({\bf k_\rho},t)= \int e^{i{\bf k_\rho}\cdot {\vec \rho}}
|\phi_{2D}(\vec \rho,t)|^2
d
\vec
\rho, 
\end{eqnarray}
 where $h_{2D}(\xi)= 2-3\sqrt \pi \xi e^{\xi^2}\mathrm{erfc}(\xi), {\bf k_\rho} \equiv (k_x,k_y)$, and the dipolar term is written in Fourier momentum space.
 
 For this study we consider $^{164}$Dy
atoms of  
  magnetic moment   $ \mu = 10\mu_B$
\cite{ExpDy} 
with 
$\mu_B$ the Bohr magneton leading to the dipolar length $a_{\mathrm {dd}}(^{164}$Dy$) \approx 132.7a_0$, 
  with $a_0$ the Bohr radius. We consider here 
 $l =1$ $\mu$m  corresponding to a radial angular trap frequency $\omega \approx  2\pi \times 
61.6$ Hz corresponding to $t_0\approx 2.6$ ms. 
 
\section{Numerical Results}

We solve the 3D equations (\ref{eq3}) and (\ref{eq4}) or the 2D equation (\ref{eq5})
by the split-step 
Crank-Nicolson discretization scheme using both real- and imaginary-time propagation
  in 3D or 2D Cartesian coordinates, respectively,
using a space step of 0.1 $\sim$ 0.2
and a time step of $0.001\sim 0.005$ in the imaginary-time simulation, and of
 $0.0004 \sim 0.002$ in real-time simulation \cite{CPC}. 
A smaller time step is employed in the real-time propagation to obtain reliable and accurate results. The dipolar potential term is treated by a Fourier transformation to the momentum space using a convolution rule \cite{Santos01}.

The stationary profile of the excited quasi-2D solitons can be obtained by imaginary-time simulation.
A symmetric initial state in the  harmonic oscillator problem converges to the ground state in imaginary-time propagation, whereas
an antisymmetric initial state converges to the lowest excited state. 
The imaginary-time routine preserves the symmetry of the initial state. 
The anisotropic 
quasi-2D soliton of  (\ref{eq3}) lies in the $y-z$ plane and we consider  three types of excited solitons: excitation in  (i)  $y$, 
 (ii)  $z$, and in both  (iii)  $y$ and $z$. 
 In imaginary-time simulation these stationary
excitations  can be obtained with an initial antisymmetric trial function (as in the harmonic oscillator problem), for example, 
$\phi({\bf r}) \sim f \exp[- x^2/2-\beta^2 y^2/2- \gamma^2 z^2/2]$,  
where $f=y$ (antisymmetric in $y$), $f=z$ (antisymmetric in $z$),  and $f=yz$ 
(antisymmetric in both $y$ and $z$)
for excitations (i), (ii), and (iii), respectively. In case of the circularly symmetric quasi-2D soliton of  (\ref{eq4}) in $x-y$ plane the excitation could be  (i) in $x$ (antisymmetric in $x$),
and in both (ii) $x$ and $y$   (antisymmetric in both $x$ $y$)
which can be obtained in imaginary-time propagation   with the initial state 
$\phi({\bf r}) \sim f \exp[-\beta^2 x^2/2 -\beta^2 y^2/2-z^2/2]$,  
where  $f=x$ and 
$xy$, respectively. 
Hence, while using the effective 2D equation (\ref{eq5}), the excited quasi-2D 
solitons of  (\ref{eq4}) can be obtained in imaginary-time 
propagation with the  initial state
$\phi({\vec \rho}) \sim f \exp[-\beta^2 x^2/2-\beta^2 y^2/2]$, where $f= x$  or $xy$ for excitations in $x$ or in both  $x$ and $y$, 
respectively. For fast convergence the constants $\beta$ and $\gamma$ in these initial 
guesses are taken to be small  
denoting large spatial extension of the excited quasi-2D soliton.

\begin{figure}[!t]

\begin{center}
\includegraphics[width=\linewidth]{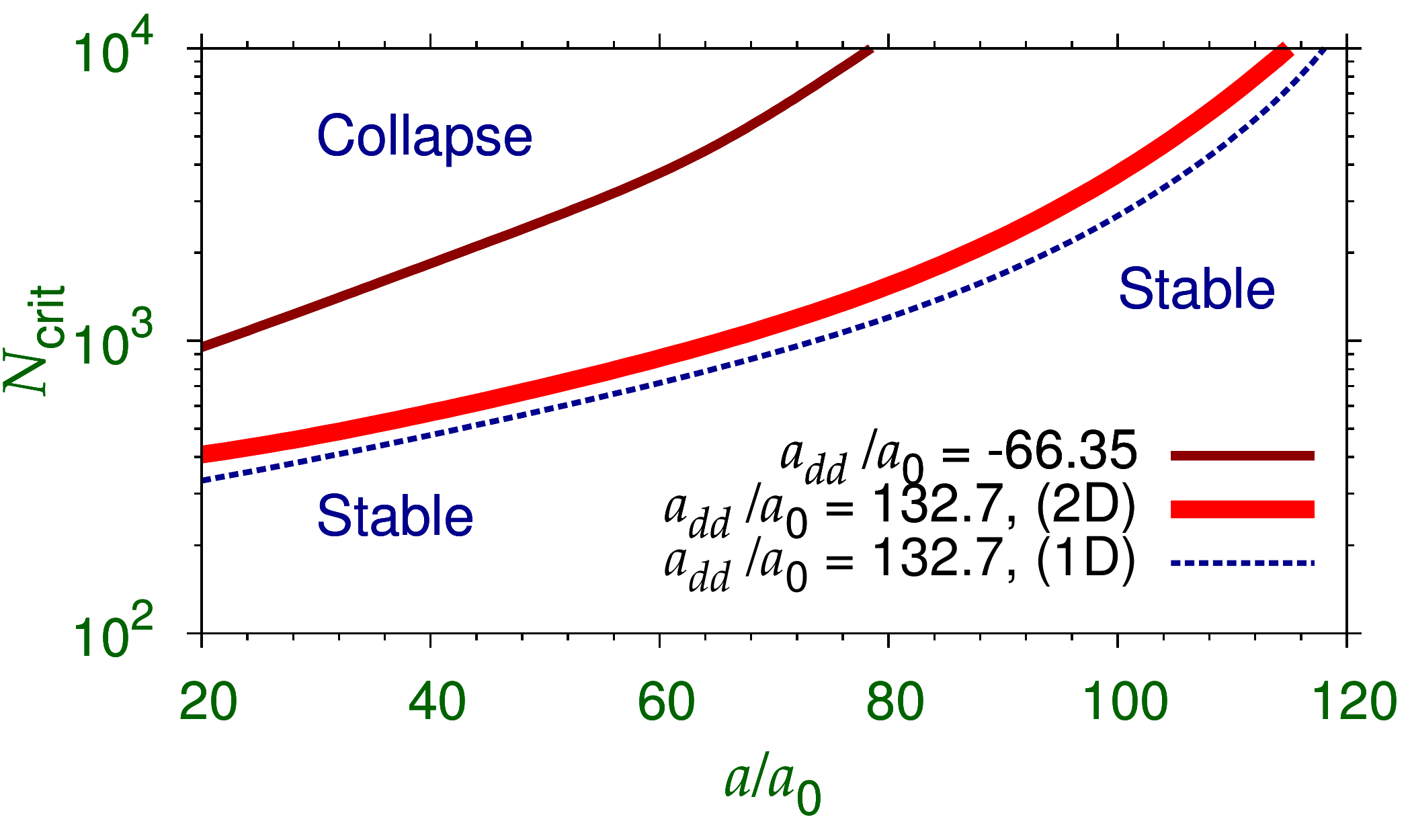}

\caption{ (Color online) Stability phase diagram showing the critical number of atoms $N_{\mathrm{crit}}$
in a  quasi-2D dipolar bright  
BEC soliton of $^{164}$Dy atoms. The results for the normal dipolar $^{164}$Dy atoms trapped   along the $x$ axis ($\alpha=1, a_{\mathrm{dd}}=132.7a_0$) as well as for the   $^{164}$Dy atoms with the dipole interaction reversed by a rotating orienting field ($\alpha=-1/2, a_{\mathrm{dd}}=-66.35a_0$)
and trapped 
along the $z$ axis are shown {in addition to the critical number of atoms in quasi-1D solitons of 
\cite{dib1d}.}
 Stable quasi-2D solitons appear for   the number of atoms $N$ below the critical number $N_{\mathrm{crit}}$. The width of the lines gives the error 
in the numerical calculation. 
The oscillator length
$l= 1 $ $\mu$m.
}\label{fig1} \end{center}

\end{figure}

In this study we consider excited quasi-2D dipolar solitons in 
a BEC of $^{164}$Dy atoms with magnetic moment $\mu = 10 \mu_B$ \cite{ExpDy}. 
The reason for considering   $^{164}$Dy atoms is that these atoms have the 
largest magnetic moment among those used in dipolar BEC experiments and a large dipole
moment is fundamental for achieving   excited quasi-2D dipolar solitons 
with a large number of atoms.   
The dipolar length in this case is $a_{\mathrm{dd}}= \mu_0\mu^2 m/(12 \pi \hbar^2)= 132.7 a_0$. 
In the case of a
sign-changed dipolar interaction by a rotating orienting field \cite{orienting}
we take $\alpha =-1/2$ and  $a_{\mathrm{dd}}= -67.35 a_0$. 
We take the harmonic trap length $l =1$ $\mu$m, corresponding  to a harmonic trap frequency of  $\omega = 2\pi \times 61.6$ Hz and time scale $t_0= 2.6 $ ms.

First we study the domain of the appearance  of the nonexcited quasi-2D solitons (ground states) of  (\ref{eq3}) and of (\ref{eq4}).
These solitons for a specific value of scattering length can exist 
for the number of atoms $N$ below a critical number $N_{\mathrm{crit}}$ beyond which the system collapses \cite{jbohn}. In figure \ref{fig1} we plot this critical 
number $N_{\mathrm{crit}}$ versus $a/a_0$ from numerical simulation for $\alpha =1, a_{dd}=132.7a_0$ and $\alpha=-1/2, a_{dd}=-66.35a_0$.    
We find that a quasi-2D soliton is possible for the   number of atoms below this critical number \cite{1D}.   In the collapse region,  the soliton collapses 
due to an excess of dipolar attraction. In the stable region there is a balance between attraction and repulsion and a stable soliton can be formed. {In \cite{dib1d} we considered the critical number of $^{164}$Dy atoms in the ground and excited states of quasi-1D solitons with a
strong trap in $x$ and $y$ directions and no trap in the $z$ direction. The critical number of 
quasi-1D ground-state solitons of \cite{dib1d} is also plotted in figure \ref{fig1} for a comparison. These two critical numbers are quite similar and the critical number for quasi-2D solitons is larger than that for quasi-1D solitons. This last finding is not unexpected in view of the stronger trap in the quasi-1D case, as compared to the quasi-2D  case, facilitating the collapse in a dipolar BEC, thus leading to a smaller critical number in the quasi-1D case.
A similar result is expected in the case of the excited quasi-2D solitons when contrasted with the quasi-1D solitons.}

The excited quasi-2D solitons to be considered next  have a larger spatial extension and can accommodate a larger number of atoms. However, due to a larger spatial extension, an accurate calculation of the critical number of atoms in excited quasi-2D solitons necessitates a huge amount of RAM  and CPU time    and the critical number of these excited solitons will not be considered here. The plots of figure \ref{fig1} represent a lower limit for the critical numbers for the excited quasi-2D solitons.
 From figure \ref{fig1} we see that the number of atoms in the quasi-2D solitons in ground and excited states  could be quite large and will be of experimental interest. 
The size of quasi-1D nondipolar solitons   is usually quite small and these solitons can accomodate only a small 
number of atoms.   A quasi-2D soliton cannot be realized in a nondipolar BEC as the long-range 
dipolar interaction plays a crucial role in their formation and stability.

\begin{figure}[!t]

\begin{center}
\includegraphics[width=.24\linewidth,clip]{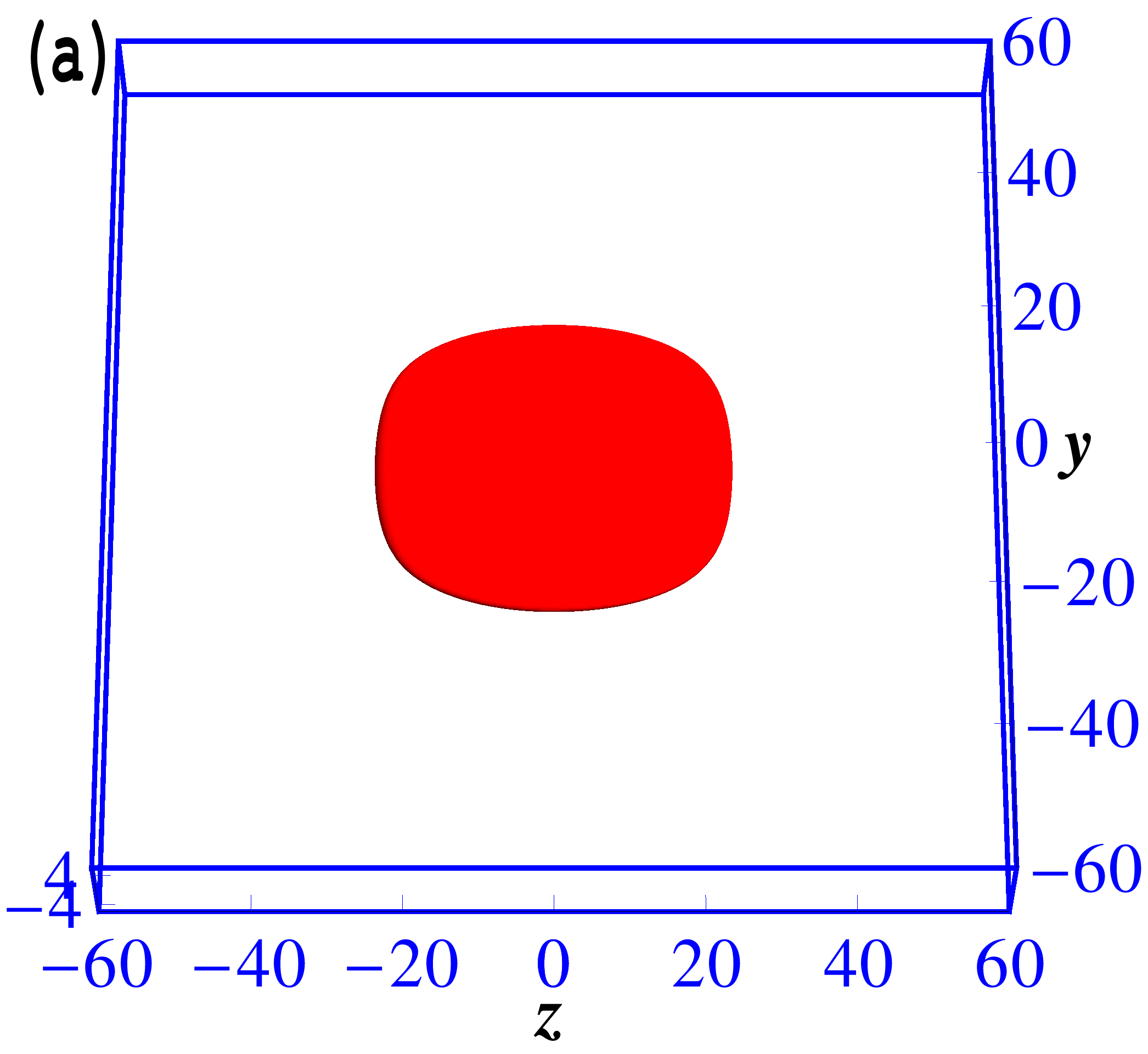}
\includegraphics[width=.24\linewidth,clip]{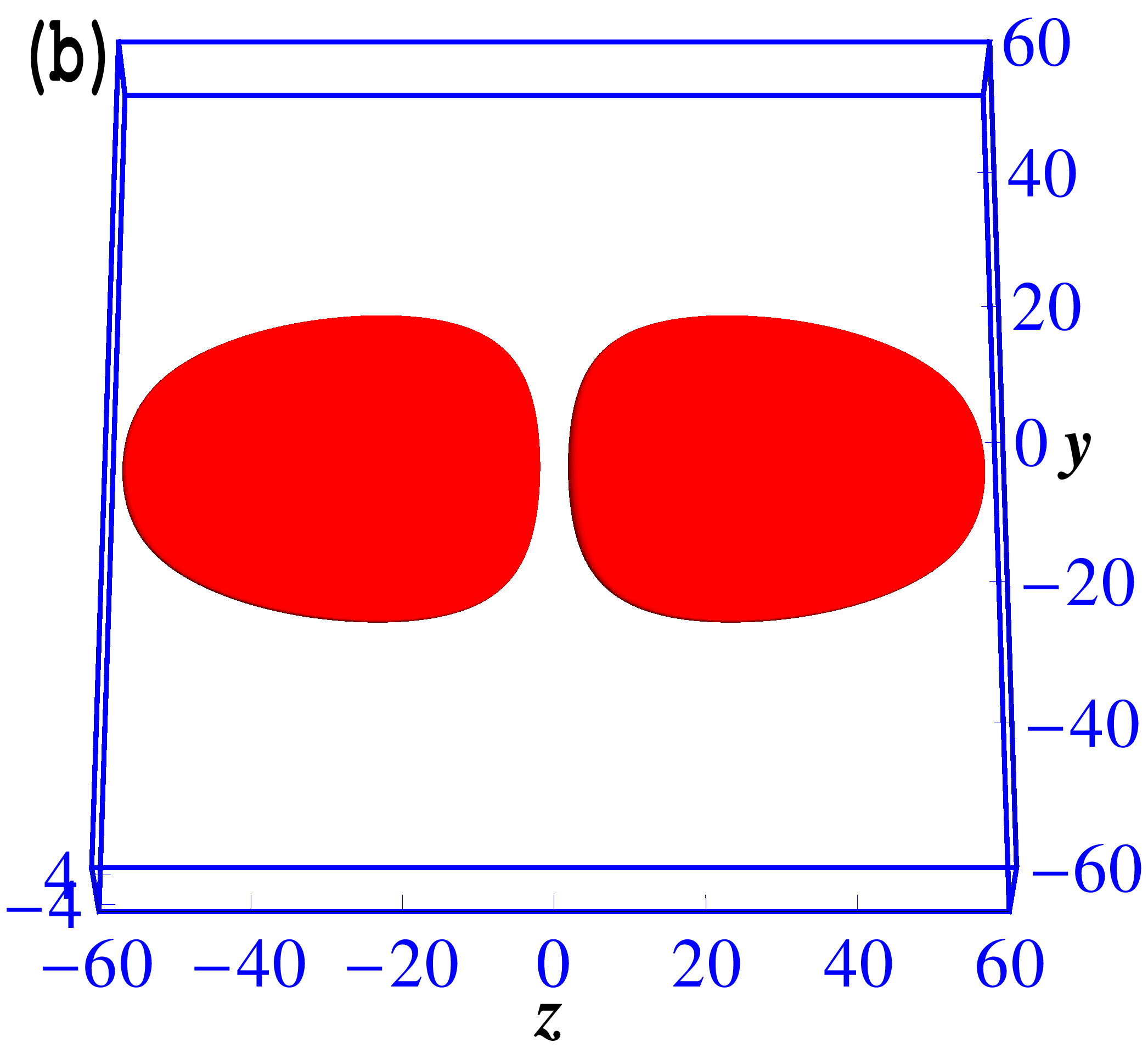}
\includegraphics[width=.24\linewidth,clip]{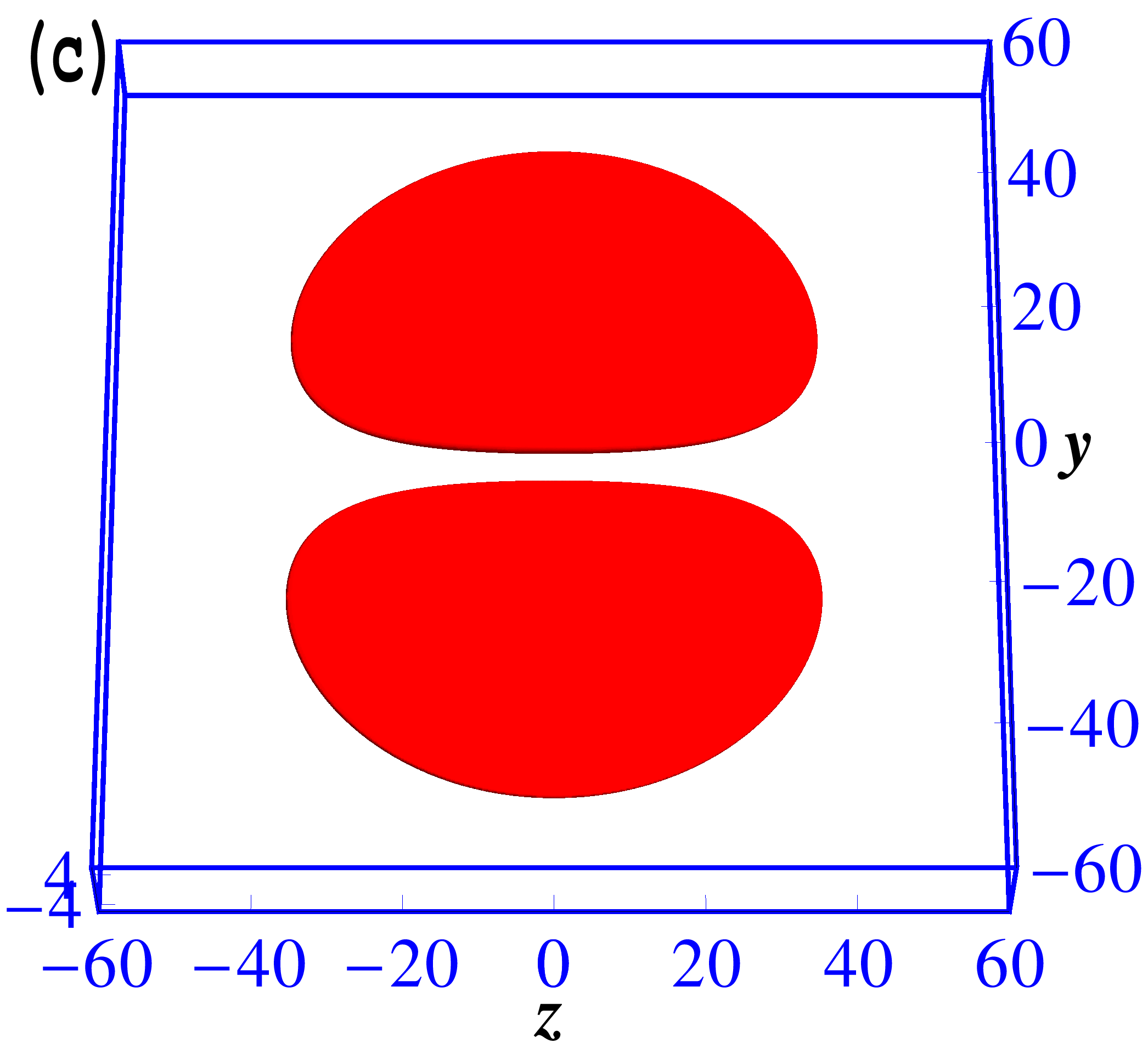}
\includegraphics[width=.24\linewidth,clip]{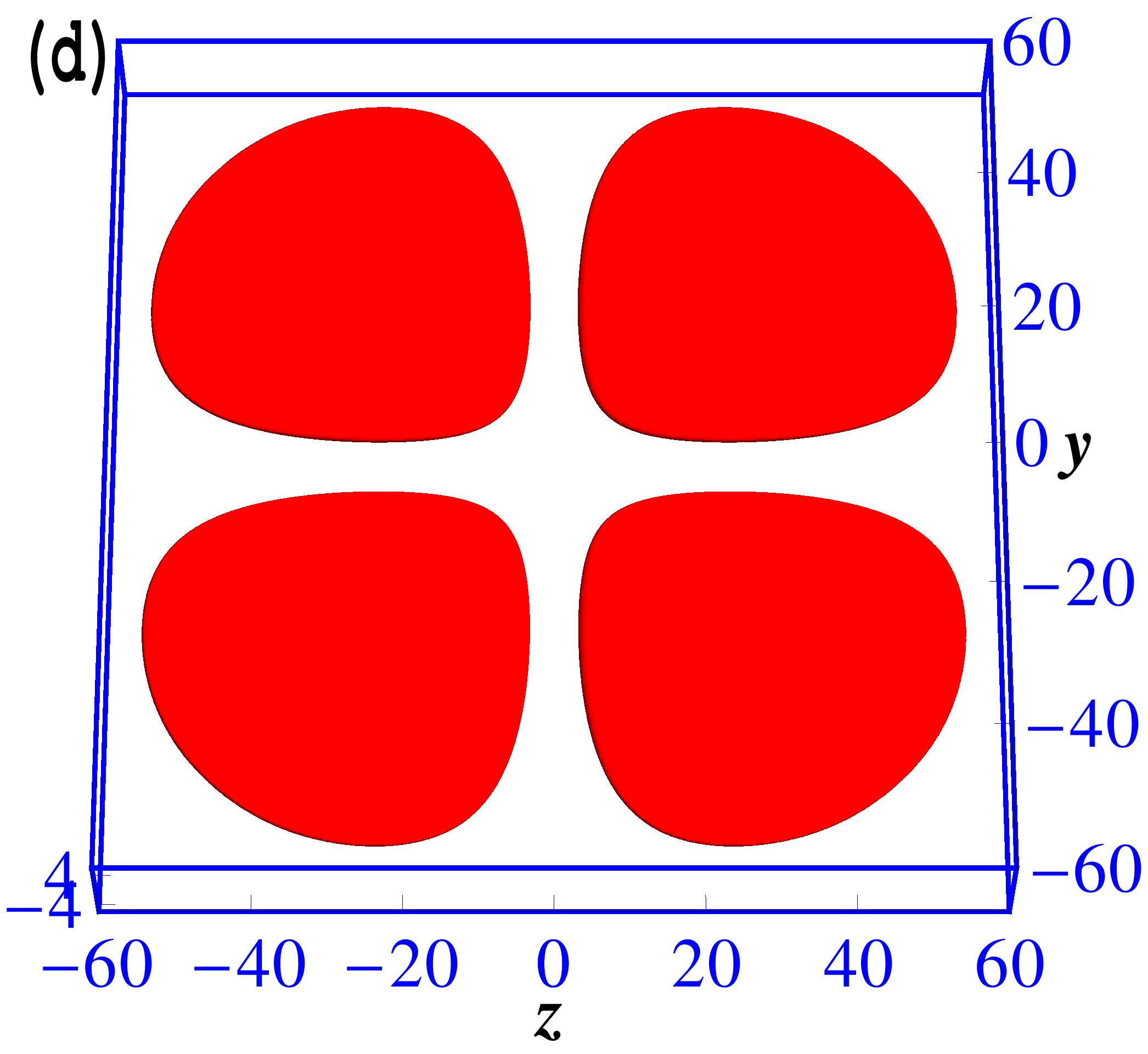}

\caption{ (Color online) (a) 3D isodensity contour ($|\phi|^2$) of a 
quasi-2D bright soliton of 1000 $^{164}$Dy atoms with  $a= 80a_0$ and $a_{\mathrm{dd}}=132.7a_0$. 
The same with one quantum of excitation in (b) $z$, (c) $y$, and in  both (d)  $y$ and $z$.
The dimensionless lengths $x,y,$ and $z$ are in units of $l$ $(\equiv 1$ $\mu$m).
The density on the contour is $ 10^{7}$ atoms/cm$^3$. 
}\label{fig2} \end{center}

\end{figure}

Next we present the profile of the quasi-2D solitons as obtained from  (\ref{eq3})
by imaginary-time propagation for 1000 $^{164}$Dy atoms with the scattering length adjusted to $a=80a_0$ by the Feshbach resonance technique \cite{fesh} and with the dipolar length $a_{dd}=132.7a_0$. A 
3D Gaussian input wave function converges to the unexcited quasi-2D
soliton illustrated in figure \ref{fig2} (a).  An input of  3D Gaussian times $z$, with a node  at $z=0$ representing an excitation in $z$, converges to the excited quasi-2D soliton of figure \ref{fig2} (b). Similarly figure \ref{fig2} (c) shows a quasi-2D soliton   with the lowest excitation in $y$ and 
figure \ref{fig2} (d) shows the same with an excitation in both  $y$ and $z$. 
From figure \ref{fig2} we find that the excitation of the quasi-2D soliton increases the spatial extension significantly for the same number (1000) of $^{164}$Dy atoms. Moreover, the excited
quasi-2D solitons can accommodate a larger number of atoms compared to the unexcited quasi-2D 
solitons implying a larger critical number $N_{\mathrm{crit}}$ shown in figure \ref{fig1}.
 The quasi-2D solitons of Figs. \ref{fig2} 
(b) and (c) with one quantum of excitation in $y$ or $z$ are larger in size compared to the ground state shown in figure \ref{fig2} (a) and the quasi-2D soliton of figure \ref{fig2} (d) with two quanta of excitation is larger   
those of Figs. (b) and (c) with one quantum of excitation each.  { The size of the 
excited quasi-2D solitons of figure \ref{fig2} is about 120 $\mu$m and it  increases rapidly 
with the scattering length $a$. The size tends to $\infty$   
as $a\to a_{\mathrm{dd}}$.}

\begin{figure}[!b]

\begin{center}
\includegraphics[width=.325\linewidth,clip]{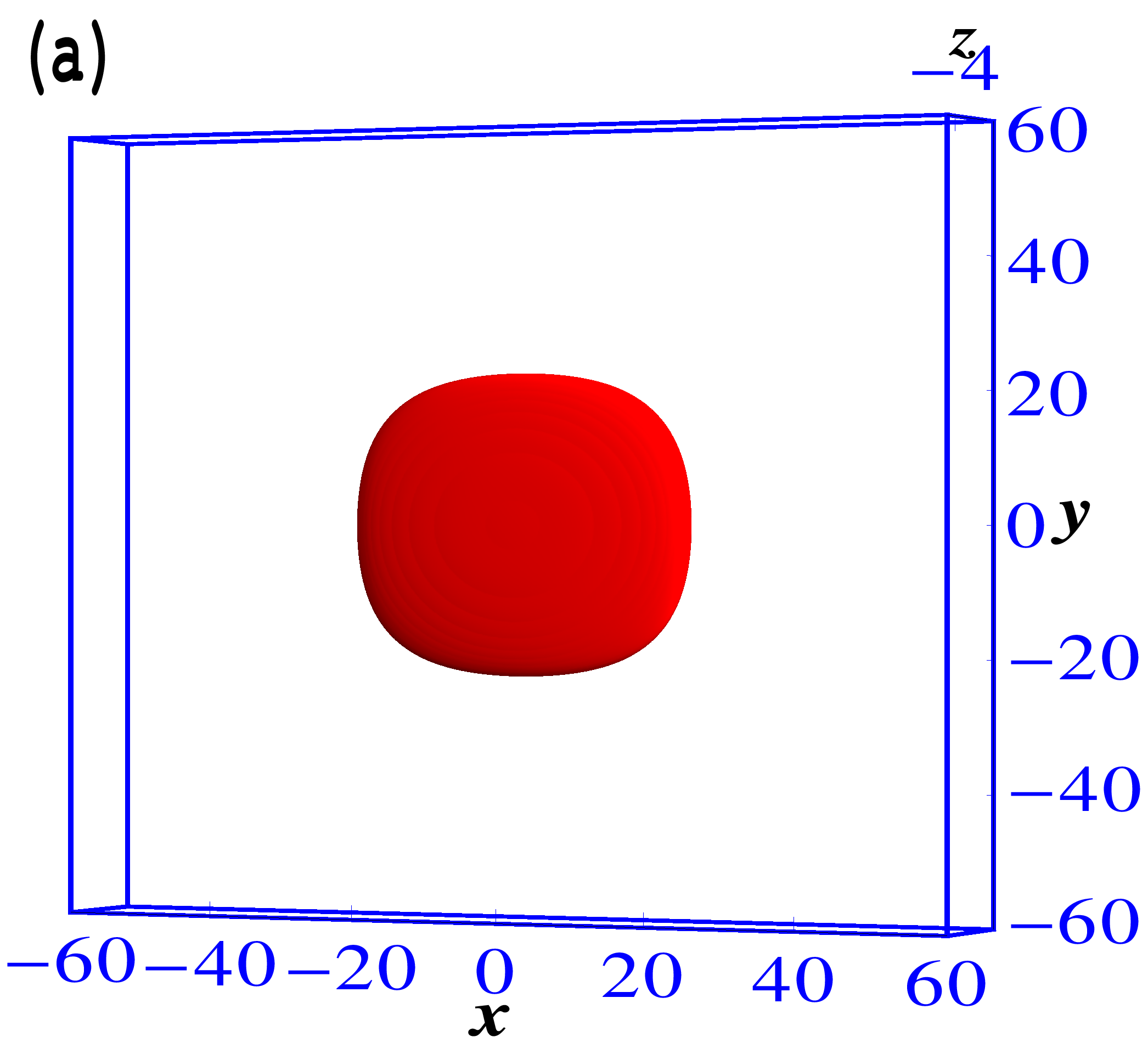}
\includegraphics[width=.325\linewidth,clip]{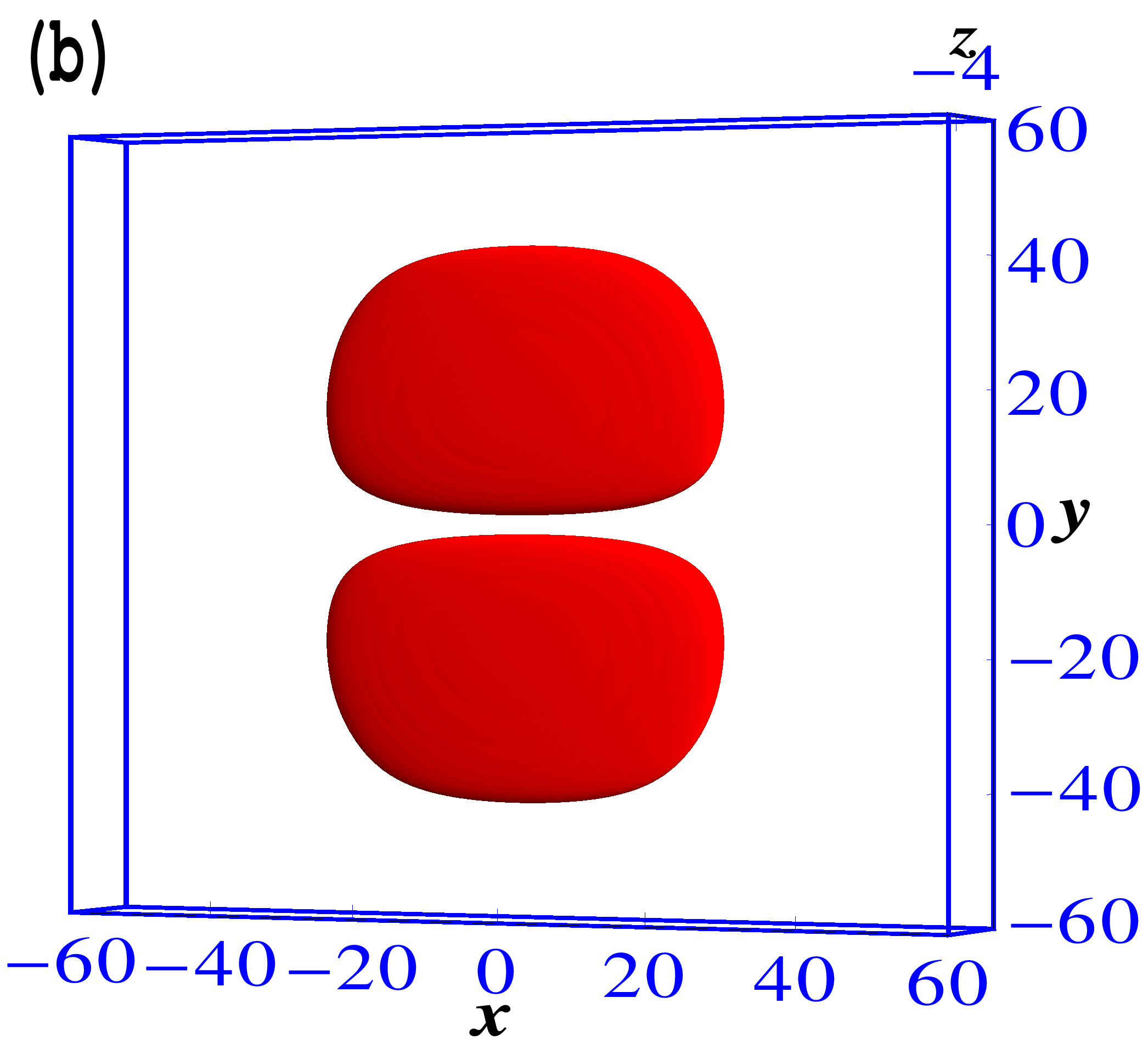}
\includegraphics[width=.325\linewidth,clip]{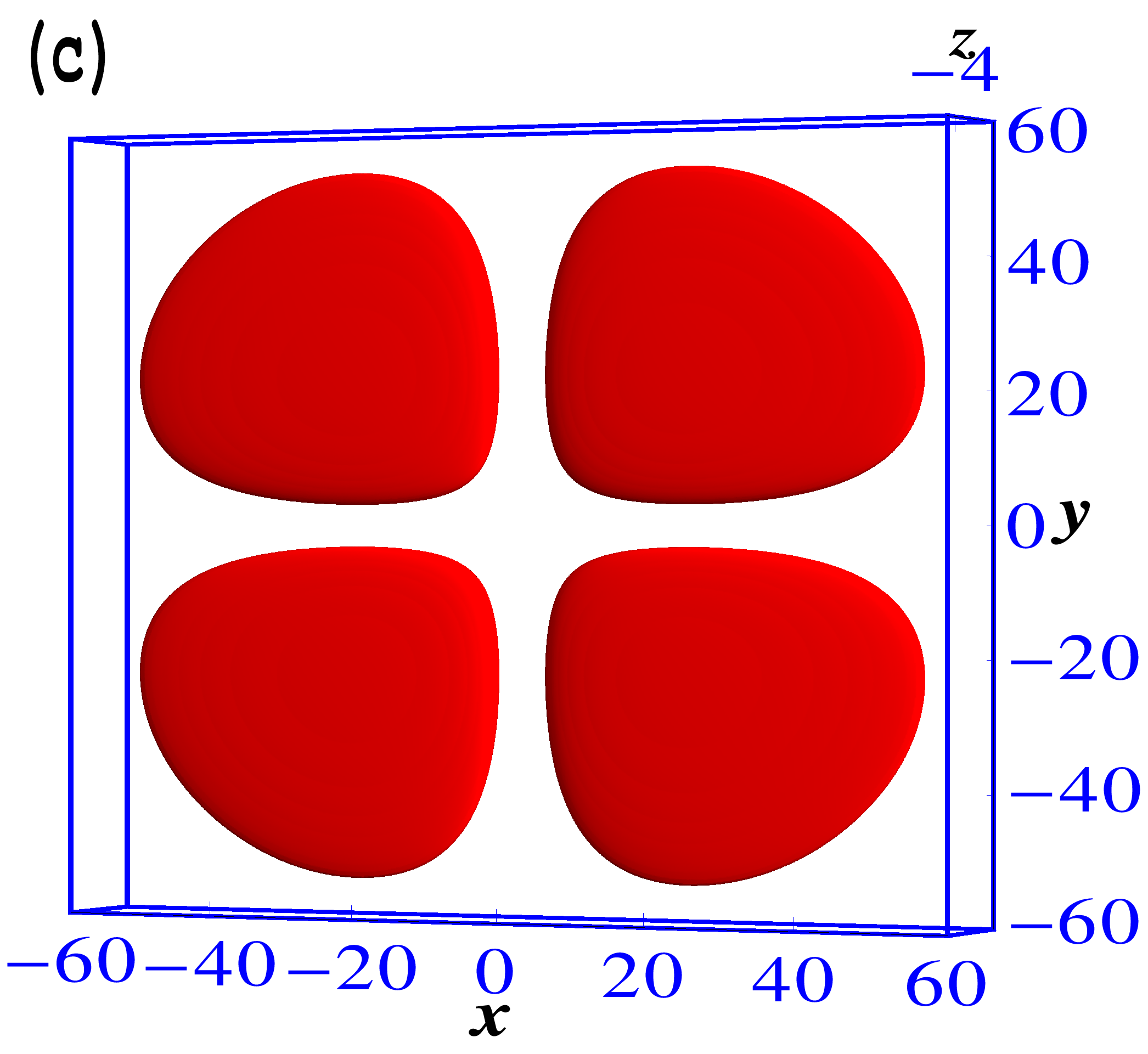}

\caption{ (Color online) (a) 3D isodensity contour ($|\phi|^2$) of a 
quasi-2D bright soliton of 1000 $^{164}$Dy atoms free to move in the $x-y$ plane 
with  sign-inverted dipolar interaction and with 
$a= 80a_0$ and $a_{dd}=-67.35a_0$. The harmonic trap is in the $z$ 
direction.
The same with one quanta of excitation in (b) $x$,  and in (c)  $x$ and $y$.
The dimensionless lengths $x,y,$ and $z$ are in units of $l$ $(\equiv 1$ $\mu$m).
The density on the contour is $ 10^{7}$ atoms/cm$^3$. 
}\label{fig3} \end{center}

\end{figure}

Similar to the Feshbach resonance technique for manipulating the scattering length 
by magnetic \cite{fesh} and optical \cite{optfesh} means, it is possible to manipulate the dipolar interaction by a rotating orienting field \cite{orienting}. Now we present results
for quasi-2D solitons with a sign-changed dipolar length of $a_{dd}=-67.35a_0$ for $^{164}$Dy atoms with an axial trap along the polarization $z$ direction 
satisfying  (\ref{eq4}). Here we consider the lowest-energy quasi-2D soliton of 1000 $^{164}$Dy atoms for the scattering length $a=80a_0$. The 3D isodensity contour is illustrated in figure \ref{fig3} (a) as obtained by imaginary-time propagation of   (\ref{eq4}).  As  (\ref{eq4}) is axially symmetric, profile of the quasi-2D soliton with one quantum of excitation in $x$ is the same as the one with one quantum of excitation in $y$ and we present the one with one quantum of excitation in $y$ in figure \ref{fig3} (b).  Finally, in figure \ref{fig3} (c) we present a quasi-2D soliton with one quantum of excitation each in $x$ and $y$. As in figure \ref{fig2}, the excited quasi-2D solitons of figure \ref{fig3}
containing the same number of atoms have larger spatial extension.  The 
size of the quasi-2D soliton of figure \ref{fig3} (c)

with two quanta of excitation is larger than that of figure  \ref{fig3} (b) with one quantum of excitation, which is larger than that of the ground state shown in figure \ref{fig3} (a).

\begin{figure}[!t]

\begin{center}
\includegraphics[width=\linewidth,clip]{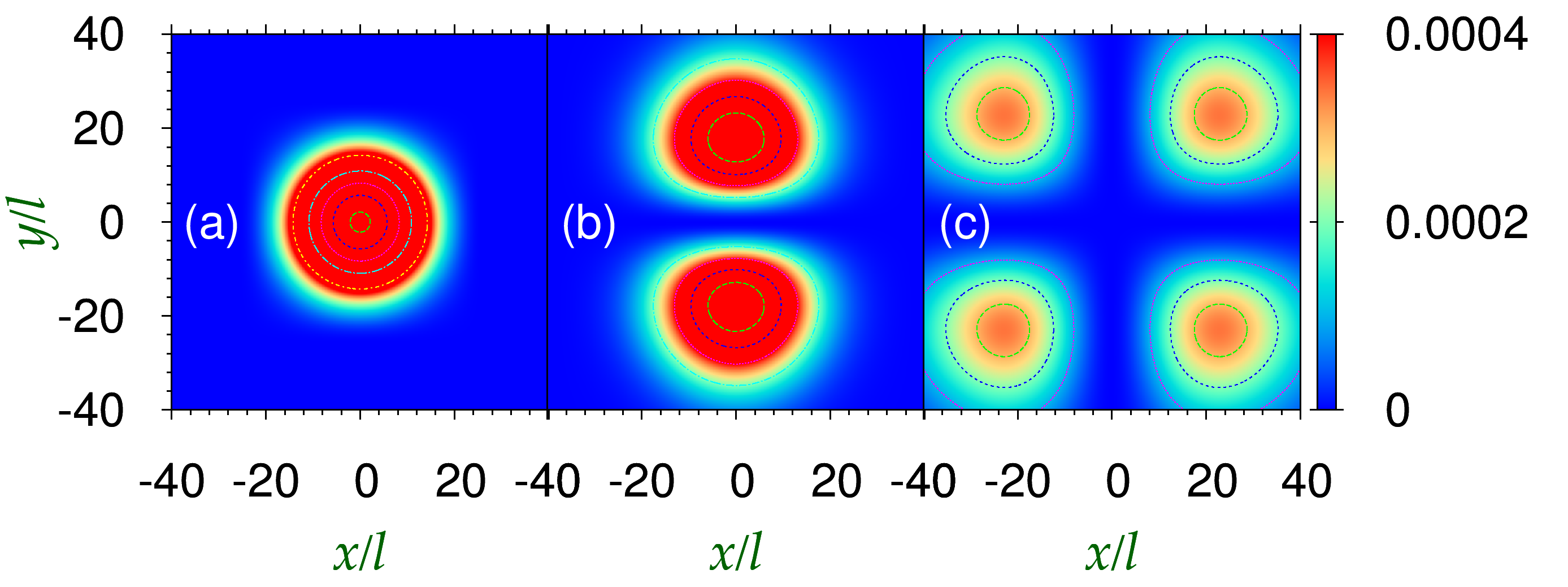}
\includegraphics[width=\linewidth,clip]{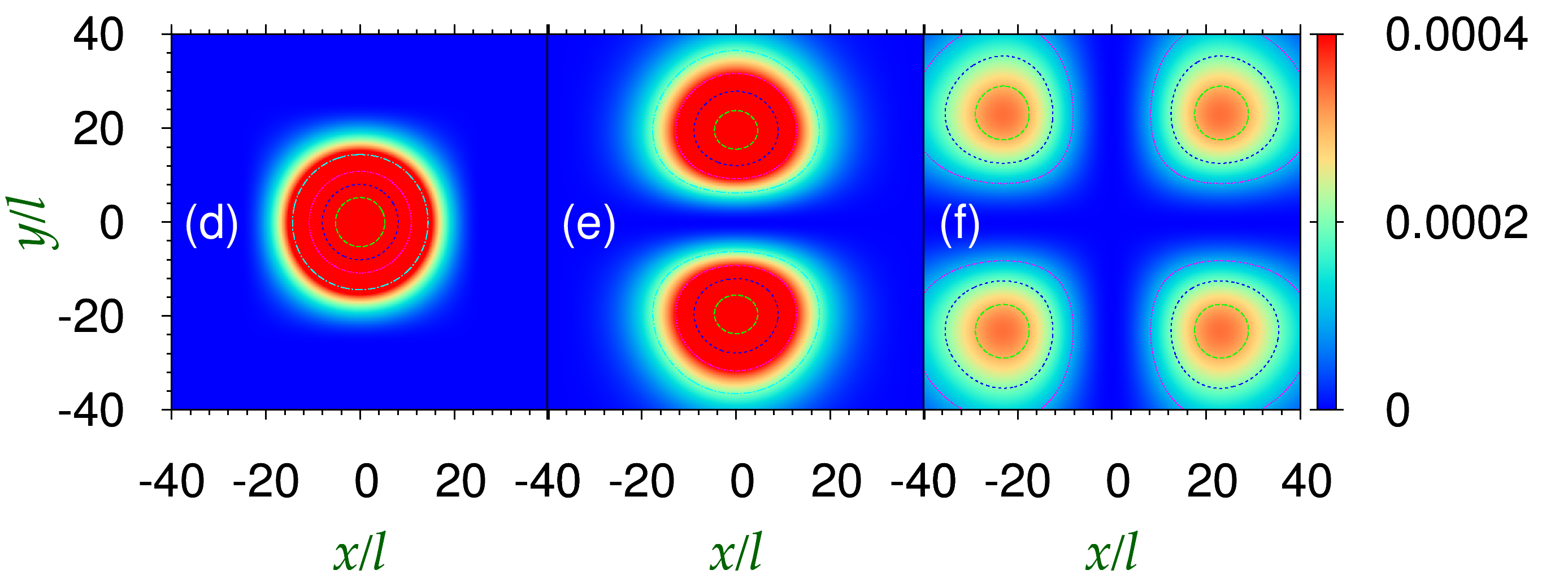}

\caption{ (Color online) (a), (b), and (c) Effective  2D contour
 ($|\phi(x,y)|^2=\int dz|\phi({\bf r})|^2$) of the  
quasi-2D  solitons of Figs. \ref{fig3} (a), (b), and (c), respectively.  (d), (e), and (f) The 2D contour plots  of 
 the  same quasi-2D solitons with the same parameters  as 
calculated using the imaginary-time solution of the effective 2D equation (\ref{eq5}).  
}\label{fig4} \end{center}

\end{figure}

The excited quasi-2D solitons presented in Figs. \ref{fig2} and \ref{fig3}
are stable and  robust as tested under real-time propagation in 3D with a reasonable perturbation in the parameters.  The robustness comes from the large contact repulsion for a reasonably large scattering length which strongly inhibits collapse. Also, an appropriate combination of the harmonic trap and long-range dipolar interaction provides confinement  of the quasi-2D soliton and prevents leakage of the atoms to infinity. In a nondipolar BEC soliton, the contact attraction alone provides the binding and there is no repulsion to stop the collapse. Consequently,   a nondipolar BEC soliton is usually fragile against collapse and auto-destruction. A stringent test of the robustness 
of these excited solitons is provided in their behavior under head-on collision. Like the quasi-1D dipolar solitons\cite{1D}, the collision of  quasi-2D solitons are expected to be quasi elastic with the solitons emerging with little deformation at medium velocities. However, at low velocities the collision is expected to be  inelastic. Only the collision between two integrable 1D solitons is known to be perfectly elastic at all velocities.

\begin{figure}[!t]

\begin{center}
\includegraphics[width=\linewidth,clip]{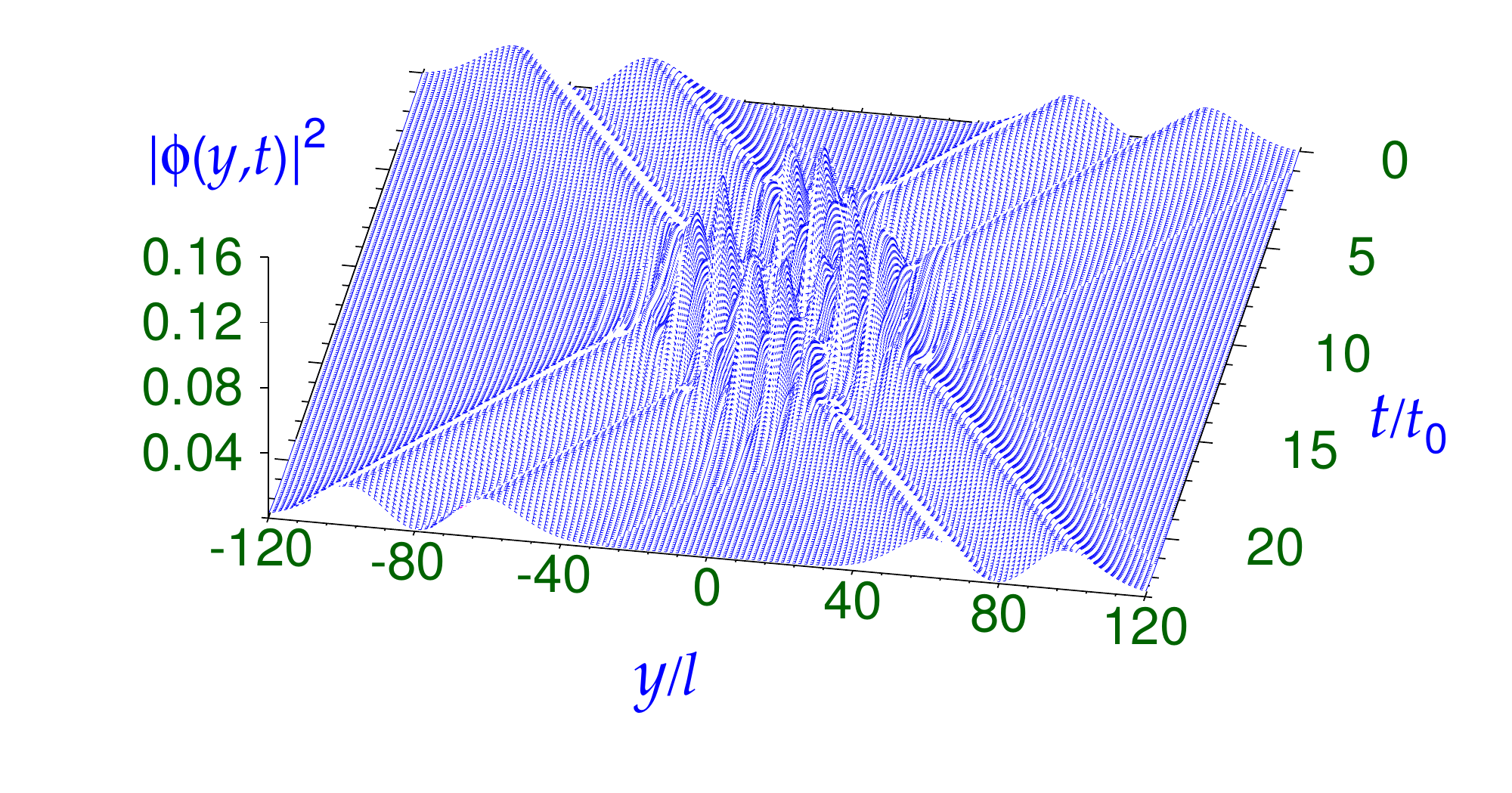}

\caption{ (Color online)  Linear  density $|\phi(y,t)|^2$ of two 
colliding  excited quasi-2D solitons of 1000 $^{164}$Dy atoms each of figure
\ref{fig3} (b) as calculated using the reduced 2D GP equation (\ref{eq5}). The  excited quasi-2D solitons free to move in the $x-y$ plane are   traveling with constant speed along the $y$ axis. 
}\label{fig5} \end{center}

\end{figure}

To demonstrate  the robustness of the solitons we consider a head-on collision between two excited quasi-2D solitons moving in opposite directions. However, because of the large spatial extension of the 
excited solitons, from a consideration of the required RAM   and CPU time, it is prohibitive to carry on these studies in 3D. Hence for the study of the dynamics of the  excited quasi-2D solitons we consider the reduced 2D equation (\ref{eq5}) in place of the 3D equation  (\ref{eq4}) with the sign changed dipolar interaction.  The reduced 2D equation (\ref{eq5}) should give a reasonable  account of the quasi-2D solitons  for medium values of contact and dipolar nonlinearity parameters.
 Before the study of this collision dynamics, we first compare the densities of a quasi-2D soliton calculated using imaginary-time propagation of the 3D and 2D equations (\ref{eq4}) and (\ref{eq5}), respectively.
 In Figs. \ref{fig4} (a), (b), and (c) we plot the effective 2D density $|\phi(x,y)|^2= \int dz |\phi(\bf r)|^2$ in the $x-y$ plane of the quasi-2D excited solitons shown in Figs. \ref{fig3} (a), (b), and (c), respectively, from a numerical solution of the 3D GP equation (\ref{eq4}). In Figs.   \ref{fig4}  (d), (e), and (f) we plot the same as obtained from the imaginary-time solution  of the  reduced 2D  (\ref{eq5}). The agreement between the two sets of densities presented in figure \ref{fig4} justifies the use of   (\ref{fig5}) for the study of the dynamics.
{
However, it should be noted that nonetheless the collision dynamics
can differ slightly in the  2D and 3D descriptions, because in the 3D case energy can be
transferred in the additional degree of freedom in the $x$ direction which is eliminated in the 
2D case \cite{eich}. }

We consider the collision dynamics between two identical excited quasi-2D solitons 
each of 1000 $^{164}$Dy atoms as illustrated in figure \ref{fig3} (b). The collision 
dynamics of two such solitons as generated from a real-time simulation of  (\ref{eq5})
is
shown in figure \ref{fig5}. The initial profiles of the two colliding solitons are obtained by a imaginary-time simulation of the same. 
The initial velocities of the two solitons  placed at $y=\pm y_0$ are attributed by multiplying the initial wave functions of the two solitons by  phase factors
$\exp (\pm iv_y y)$. The parameter $v_y$ is chosen by trial. The two excited quasi-2D solitons are initially placed at $y=\pm 80$ and the real-time simulation started. The two solitons move in opposite directions and suffer a head-on collision. The collision dynamics is best illustrated by plotting  
effective linear density along $y$ direction $|\phi(y,t)|^2 = \int dx |\phi_{2D}(\vec \rho,t)|^2$ versus $y$ and $t$ as shown in figure \ref{fig5}. The velocity of each   soliton is about $2.67$ mm/s using $l_0=1$ $\mu$m and $t_0=2.6$ ms. The smooth density profiles of the dynamics presented in figure \ref{fig5} illustrates the quasi elastic nature of the dynamics.

\begin{figure}[!t]

\begin{center}
\includegraphics[width=\linewidth,clip]{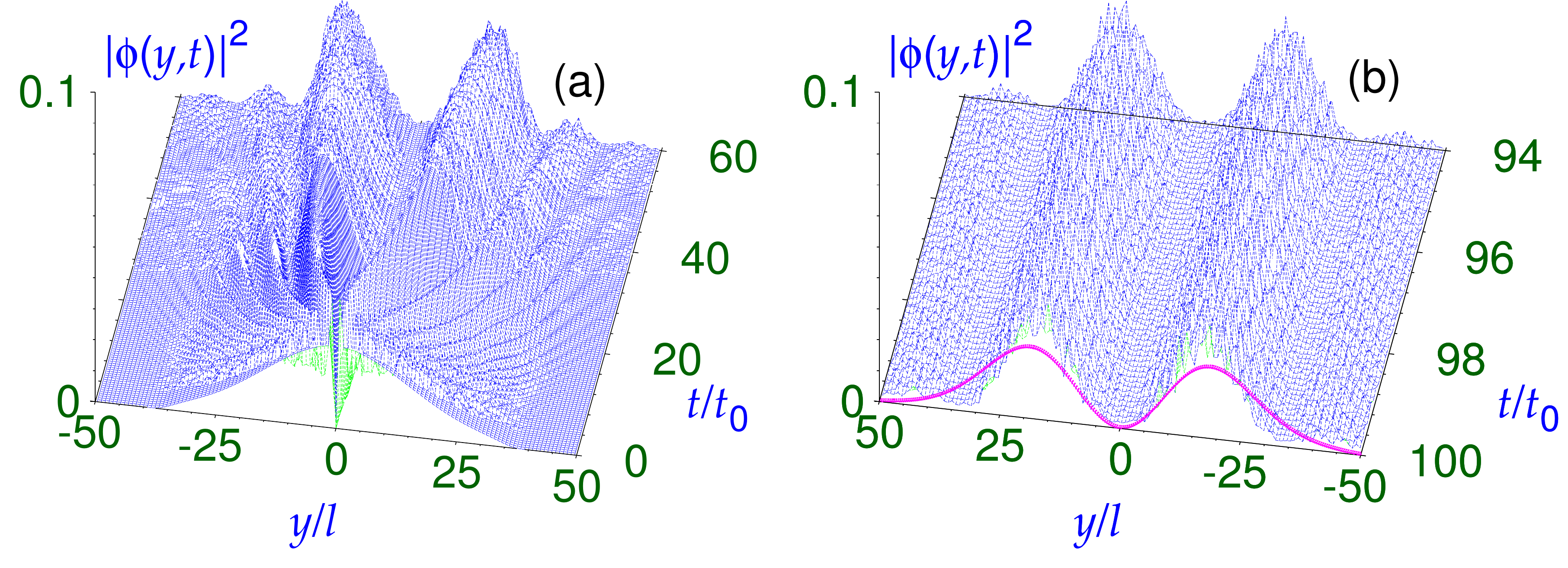}

\caption{ (Color online) Creating the excited quasi-2D soliton of figure \ref{fig3} (b)  by real-time simulation of the reduced 2D equation (\ref{eq5}) with the 
phase imprinted quasi-2D soliton  of figure \ref{fig3} (a). Linear density of the excited quasi-2D soliton (blue) for (a) small and (b) large times. In (b) the constant density of the stationary excited state (red) is shown for comparison.  
}\label{fig6} \end{center}

\end{figure}

As the excited quasi-2D solitons are stable and robust, they can be prepared by phase imprinting \cite{phase} a bright soliton.
In experiment a homogeneous potential generated by the dipole potential of   a far detuned 
laser beam is applied on one half of the bright soliton ($y<0$) for an interval of time so as to imprint an extra phase of $\pi$ on the wave function for $y<0$ \cite{darksol}. The thus phase-imprinted  quasi-2D bright soliton is propagated in real-time, while it slowly transforms into a excited quasi-2D soliton. This simulation is done with no axial trap. In actual experiment a very weak axial trap can be kept during generating 
the excited quasi-2D  soliton, which can be  eventually removed.  The simulation is illustrated in figure \ref{fig6}, where we plot the linear axial density versus time. It is demonstrated that at large times the linear density tends towards that of the stable   excited quasi-2D soliton.

\section{Summary }

We demonstrated  the possibility of creating mobile, stable,  excited quasi-2D solitons in dipolar BEC with a notch in the central plane and capable of moving in a plane with  a constant velocity. These solitons are stationary solutions of the mean-field GP equation.
The head-on collision between two such solitons with a relative velocity of about 
5 mm/s is quasi elastic with the solitons passing through each other with practically no deformation. A possible way of preparing these excited quasi-2D  solitons by phase imprinting a bright soliton 
is demonstrated using real-time propagation in a  mean-field model.   The results and conclusions  of this paper can be  tested in experiments with present-day know-how and 
technology  and should lead to interesting future investigations.

We thank  
FAPESP  and  CNPq (Brazil)  for partial support.

\end{document}